\documentclass[preprint, journal]{vgtc}                     

\onlineid{0}

\vgtccategory{Research}

\vgtcpapertype{please specify}

\title{Visual Support for the Loop Grafting Workflow on Proteins}

\author{
\authororcid{Filip Op\'alen\'y}{0000-0001-9438-396X},
\authororcid{Pavol Ulbrich}{0000-0003-1661-7905},
\authororcid{Joan Planas-Iglesias}{0000-0002-6279-2483}, 
\authororcid{Jan By\v{s}ka}{0000-0001-9483-7562}, 
\authororcid{Jan \v{S}toura\v{c}}{0000-0003-3139-3700},\\
\authororcid{David Bedn\'{a}\v{r}}{0000-0002-6803-0340}, \authororcid{Katar\'{i}na Furmanov\'{a}}{0000-0003-2805-8784}, \authororcid{Barbora Kozl\'ikov\'a}{0000-0003-0045-0872}}

\authorfooter{
\item
 F. Op\'alen\'y, P. Ulbrich, K. Furmanov\'{a} are with the Department of Visual Computing, Faculty of Informatics, Masaryk University, Brno, Czech Republic.
\item
 J. By\v{s}ka is with the Department of Visual Computing, Faculty of Informatics, Masaryk University, Brno, Czech Republic, and Department of Informatics, University of Bergen, Bergen, Norway.
\item
 J. Planas-Iglesias, J. \v{S}toura\v{c}, and D. Bedn\'{a}\v{r} are with Loschmidt Laboratories, Department of Experimental Biology and RECETOX, Faculty of Science, Masaryk University, Brno and with International Clinical Research Center, St. Anne’s University Hospital Brno, Czech Republic
 \item
 B. Kozl\'ikov\'a is with the Department of Visual Computing, Faculty of Informatics, Masaryk University, Brno, Czech Republic. E-mail: kozlikova@fi.muni.cz
}

\abstract{%
In understanding and redesigning the function of proteins in modern biochemistry, protein engineers are increasingly focusing on exploring regions in proteins called loops. Analyzing various characteristics of these regions helps the experts \rev{design} the transfer of the desired function from one protein to another. This process is denoted as loop grafting. We designed a set of interactive visualizations that provide experts with visual support through all the loop grafting pipeline steps. The workflow is divided into several phases, reflecting the steps of the pipeline.
Each phase is supported by a specific set of abstracted 2D visual representations of proteins and their loops that are interactively linked with the 3D View \rev{of} proteins. By sequentially passing through the individual phases, the user \rev{shapes} the list of loops that are potential candidates for loop grafting. Finally, the actual in-silico insertion of the loop candidates from one protein to the other is performed, and the results are visually presented to the user. In this way, the fully computational rational design of proteins and their loops results in newly designed protein structures that can be further assembled and tested through in-vitro experiments. 
We showcase the contribution of our visual support design on a real case scenario changing the enantiomer selectivity of the engineered enzyme. Moreover, we provide the readers with the experts' feedback.
}

\keywords{Protein visualization, protein engineering, loop grafting, abstract views}

\teaser{
  \centering
  \includegraphics[width=\linewidth]{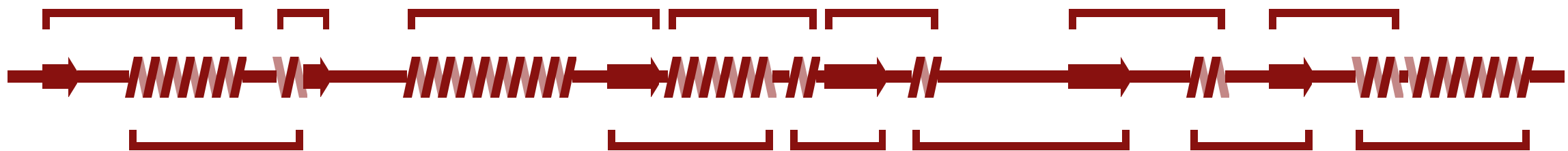}
  \caption{\label{fig:teaser}
           Linear representation of a protein (PRB ID 1isp) and it's loops.}
           
}

\graphicspath{{figs/}{figures/}{pictures/}{images/}{./}} 

\usepackage{tabu}                      
\usepackage{booktabs}                  
\usepackage{lipsum}                    
\usepackage{mwe}                       
\usepackage{textgreek}
\usepackage{fontawesome}
\usepackage{url}
\usepackage{xspace}
\usepackage{enumitem}
\usepackage{mathptmx}                  

\newcounter{pipelinestepc}

\newcommand{\pipelinestep}{\stepcounter{pipelinestepc}step~\arabic{pipelinestepc}}
\renewcommand{\texttheta}{\straighttheta}

\newcommand{\kiraaRemove}[1]{}
\newcommand{\rev}[1]{{\color{black}{#1}}}
\newcommand{\revtwo}[1]{{\color{black}{#1}}}
\newcommand{\green}[1]{{\color{black}{#1}}}

\newcommand{\ie}{i.e.,~}
\newcommand{\eg}{e.g.,~}

\newcommand{\revca}[1]{{\color{black}{#1}}}

\begin{document}


\firstsection{Introduction}
\maketitle

\begin{figure}[htb]
  \centering
  \includegraphics[width=\linewidth]{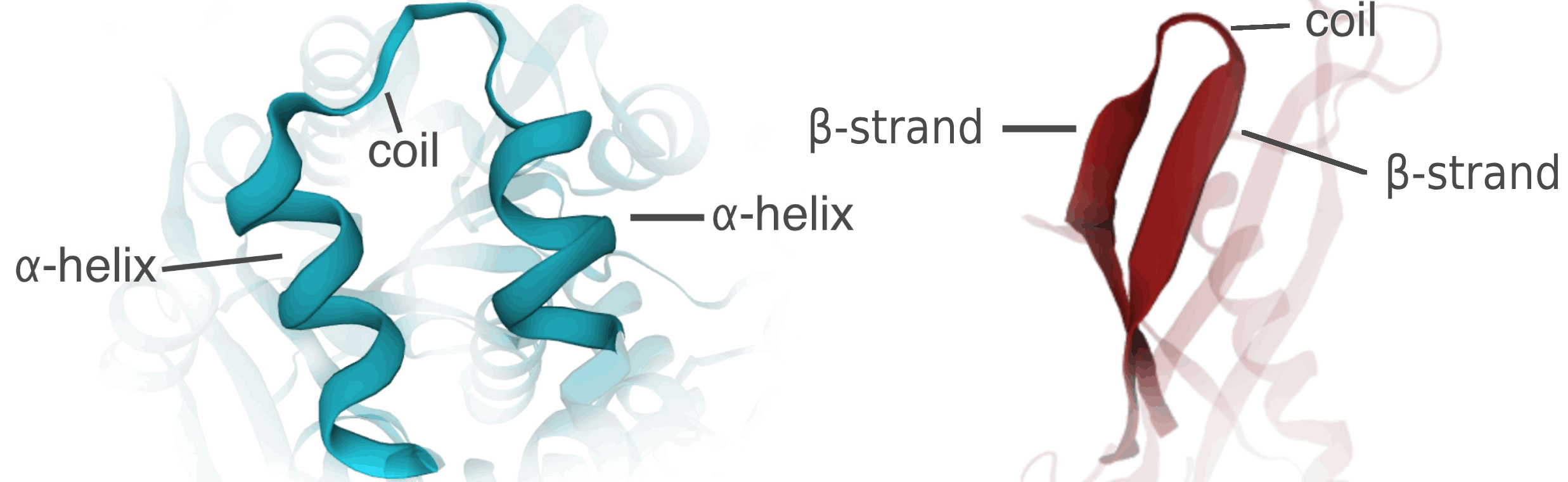}

  \caption{On the left, a loop consisting of two $\alpha$-helices and a coil. On the right, a loop consisting of two $\beta$-strands and a coil.}

    \label{fig:loop}     

\end{figure}

Protein engineering aims to improve different properties of natural biocatalysts (i.e., enzymes) that have evolved over thousands of millions of years to perform complex chemical reactions required to sustain life. 
Proteins consist of sequences of amino acids arranged in the three-dimensional space \rev{following either periodic or aperiodic patterns called secondary structures}. 
Among the basic periodic secondary structures belong \textit{$\alpha$-helices} and \textit{$\beta$-sheets} \rev{composed of individual $\beta$-strands}.
A region \rev{without any} periodic secondary structure is referred to as aperiodic and is often denoted as a \textit{coil}.
By understanding protein structure and its properties, protein engineers can design replacements of amino acids that lead to the desired change in protein behavior.
Recent design efforts have specifically targeted so-called \textit{loops} in proteins.
\revca{In structural biology, the term loop is typically used to refer to an aperiodic region of the protein (here referred to as coil). However, for the purposes of geometrical classification of loops, we use an extended definition that includes the aperiodic region and two periodic secondary structures enclosing it (see Figure~\ref{fig:loop}).} 
Loops are often highly chemically active parts of the protein, crucial for its function, such as ligand binding, protease inhibition, or protein-protein interactions \cite{espadaler_querol_aviles_oliva_2006}.

A particularly challenging task in \rev{protein engineering} is transferring the desired property between two proteins by means of \textit{loop grafting} \rev{(\ie by exchanging one loop for another)}. 
To aid the selection of grafted loops, our collaborators from the protein engineering field, \rev{who are} also co-authoring this paper, designed the loop grafting pipeline \cite{schenkmayerova2020engineering}. 
A successful loop transfer requires a certain degree of geometric overlay of \rev{original} loops with their replacement~\cite{fragrus}. 
Furthermore, multiple loops contributing to the same function \rev{must} be considered during the loop grafting.
The decisions to keep or replace a loop are also driven by 
\rev{numerous physico-chemical properties}.
\rev{The multitude of frequently conflicting selection parameters makes fully automatic solutions unfeasible. On the other hand, the complexity of the information
makes it difficult to analyze the loops without proper visual support.}

\begin{figure*}[t]
  \centering
  \includegraphics[width=\textwidth]{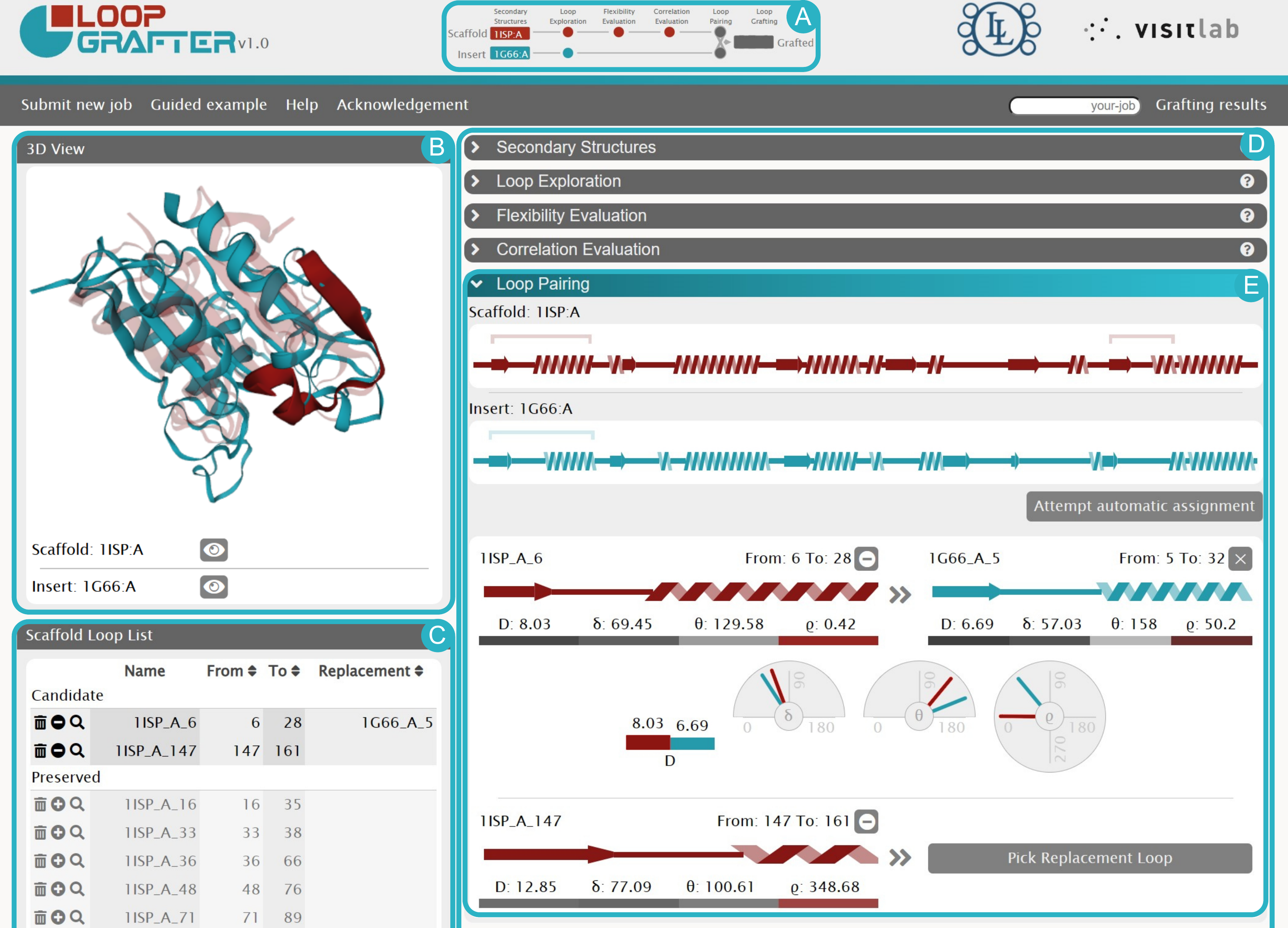}
  \caption{\label{fig:overview}
            \rev{Proposed visualization layout}. 
            (A) The overview \rev{that} serves as the navigation through the loop grafting pipeline. (B) 3D View \rev{showing} a superimposition of the two proteins, with loop 1ISP\_A\_147 highlighted \rev{(red)}. \rev{(C) Scaffold Loop List showing identified candidate (top) and the unprocessed (bottom) loops. The} list is enriched with contextual information \rev{based on the current state of the pipeline} (now showing selected \rev{replacements} for candidate loops). 
            (D) Each grafting pipeline phase is represented by a single collapsible tab that contains specific visual representations and interactions. \rev{For example, the Loop Pairing phase (E)} consists of 1D representations of proteins and their loops. Underneath, custom visualizations allow users to compare individual loops and pick desired replacements. 
           }
\end{figure*}

\rev{Unfortunately,} the existing visual representations of protein secondary structures are not targeting loops, their properties, and their mutual interplay.
Therefore, in this project, our interdisciplinary team of experts in protein engineering and visualization designed a set of interactive abstract visual representations (Figure~\ref{fig:overview}) supporting the protein loop grafting pipeline~\cite{schenkmayerova2020engineering}. 


\rev{In our research, we adopted a design study methodology~\cite{sedlmair2012design}. We analyzed the current workflow employed by protein engineers within the grafting pipeline, identified weak points, and derived a set of requirements to execute the grafting pipeline successfully. Following these requirements, we designed suitable visual representations and validated the prototype implementation by applying it to a real-world scenario.}
The abstract visual representations herein presented provide the experts with intuitive support for their workflow, give them the necessary insight into the loop properties, and significantly increase the chance of designing biochemically relevant proteins with the desired function. \rev{Moreover, since loop grafting is a niche and relatively novel area of protein engineering, the visually guided solution can also inspire the trust of the novice biologist into the process~\cite{dasgupta2016familiarity}.}

\rev{Our solution was implemented in a web-based application called LoopGrafter~\cite{loopgrafter-nar2022}.
In this paper, we present the rationale behind the design of the individual visualizations \rev{and discuss possible alternatives}. We also explain how the visualization itself can be used to support complex pipelines. 
For the biological rationale behind the loop grafting process, the use of the LoopGrafter tool itself, or its integration with bioinformatical tooling \cite{das2008macromolecular, shen2006statistical}, we kindly refer the reader to the publication by Planas-Iglesias et al. \cite{loopgrafter-nar2022}. In summary, the main contributions of \rev{this design study} are the following:
\begin{itemize}

\item Process of abstraction for the complex loop grafting pipeline by grouping semantically related tasks into logical units to provide intuitive navigation and avoid attention fragmentation.
\item Design of supporting visual representations by use of established visualization concepts and consistency across the entire workflow to alleviate the cognitive load.

    \item Demonstration of the usefulness of the proposed visual representations in a real-case scenario.
\end{itemize}
}
\section{Background}
\label{sec:background}
Loop grafting is a complex process of protein design based on transferring loop regions that \rev{provide} desired functionality from the source protein \rev{(called \textit{Insert})} to the target protein \rev{(called \textit{Scaffold})}. 
\rev{These proteins typically possess very similar spatial structures but exhibit substantially different properties.}
The design of loop grafting consists of several steps, formalized by our collaborating protein engineering experts into so-called \textit{loop grafting pipeline} \cite{schenkmayerova2020engineering}. 
The loop grafting pipeline is a methodical way of approaching the loop grafting problem, where the protein engineer's goal is to \rev{identify several so-called \textit{candidate loops} that could be verified in the wet lab.}
\rev{However, protein engineers face a dilemma in which selecting a suitable loop based on one property may lead to detriment in another property. Consequently, automatically finding a loop excelling in all properties is often infeasible, necessitating an expert intervention to find a compromise.}

The loop grafting pipeline consists of \rev{twelve} steps.
\revca{To this date, the process was mostly manual and involved working with multiple tools.}
The selection of \rev{Scaffold and Insert} proteins (\pipelinestep) \revca{includes aligning and visualizing the proteins in 3D which is typically done with PyMOL~\cite{delano2002pymol}.}
\revca{The next phase includes} the assignment of \revca{protein} secondary structures (\pipelinestep). This is typically calculated using the DSSP algorithm~\cite{DSSP}.
\rev{Nevertheless, adjustment based on expert knowledge (\pipelinestep) may be required to improve the accuracy of the following steps.} \revca{This would necessitate manual editing of the corresponding DSSP result files.} 
\rev{Afterward, the protein loops are computed, and their} geometric properties are \rev{extracted} based on the 3D arrangement of secondary structures (\pipelinestep). \revca{Both ArchDB~\cite{archdb} or Frag R’ Us~\cite{fragrus} can be used for this purpose. The first is a database that provides results on the closest recorded structure, which would require manual recording for its posterior use. The second tool performs actual calculations on input structures but produces a number of results files that require dedicated parsing.}
\rev{In the next step, experts must explore these spatial properties to identify suitable and discard unsuitable} loops for future grafting (\pipelinestep).
\rev{Furthermore,} to detect chemically active regions, the flexibility is computed for the amino acids, secondary structures, and loops (\pipelinestep).
\rev{Flexibility values can be directly obtained from crystallographic structures as experimental temperature factors (stored in the PDB files) or computed in silico. The in silico methods include computationally inexpensive elastic network models (e.g., Anisotropic Network Model (ANM)~\cite{anm} and Gaussian Network Model (GNM)~\cite{gnm}) or more precise (but computationally expensive) molecular dynamics simulations (MD)~\cite{planas2021computational}.
\revca{The elastic network models can be calculated using ProDy~\cite{zhang2021prody}, but again, its results need parsing, mapping to secondary structure elements, and re-calculation.}
These methods typically describe the flexibility of each protein amino acid, although there may be some variability in the values, requiring manual assessment}.
Based on the flexibility analysis, \rev{experts can further narrow down a} set of the possible loops \rev{for grafting} (\pipelinestep). \rev{In the next step, the} motion cross-correlation analysis with all remaining loops of the Scaffold protein is performed (\pipelinestep). 
\rev{Using these results and their prior knowledge, the experts can select loops exhibiting similar motion (and thus potentially similar function) with already identified loops and add them} to the pool of grafting candidates (\pipelinestep). 
\rev{Then, the experts must decide how to} pair the \rev{Scaffold} candidate loops with their replacements from the Insert protein based on their geometrical properties (\pipelinestep). \revca{To the best of our knowledge, no tool currently exists to support users in this step. As a result, experts must manually process all the complex data from the previous steps.}
The actual replacement of loops is performed using modeling tools, such as Modeller~\cite{modeller} or Rosetta~\cite{das2008macromolecular}. \rev{However,} these tools can produce thousands of possible solutions that vary in the start and end point of loop replacement and the geometry of the final structures.
\rev{Therefore,} the resulting grafted protein models are obtained along with their DOPE~\cite{shen2006statistical} and Rosetta~\cite{das2008macromolecular} scores (\pipelinestep). \rev{The experts then \revca{manually} inspect the placement and scores of the results \revca{(typically done with PyMOL~\cite{delano2002pymol} and some spreadsheet application)} and select a required} number \revca{of} models for consequent in-vitro experiments (\pipelinestep).

\revca{As can be seen above,} several steps of the loop grafting pipeline \rev{produce} complex numerical outputs that \rev{must} be correctly interpreted and passed to the following steps. Appropriate support of the automatic chaining of these steps and, more importantly, proper visual representations of outcomes of these steps are crucial for the actual usage of the pipeline and for aiding the experts in the decision-making process.

\section{Related Work}
\rev{This section first discusses} general molecular visualization techniques and systems most closely related to our work. For a comprehensive overview of the existing techniques, we kindly refer the readers to the survey published by Kozl\'{i}kov\'{a} et al.~\cite{KozlikovaSTAR}. Then, we will discuss the techniques specifically designed for the visualization of protein flexibility and cross-correlation, the two techniques used in the loop grafting pipeline. Finally, we will present the existing systems dedicated to protein loop engineering.

\subsection{Molecular Visualizations}
There are several tools, such as PyMOL~\cite{delano2002pymol}, VMD~\cite{humphrey1996vmd}, CAVER Analyst~\cite{caveranalyst}, YASARA~\cite{krieger2014}, USCF Chimera~\cite{pettersen2004ucsf}, or \rev{Mol*~\cite{sehnal2018mol}} that provide the users with \rev{multitude} of mostly spatial \rev{representations of molecular data, with limited range of additional abstract visualizations.}  
\rev{However}, the spatial representation is unnecessary for some tasks, or it may even hinder them due to the visual clutter. For instance, biologists often need to compare larger numbers of \green{molecules based on their structural properties, but superimposing the molecules would suffer from visual clutter. To overcome this problem, tools like MedChemLens~\cite{shi2022medchemlens} or ChemVA~\cite{sabando2020chemva} utilize simplified representations such as SMILES~\cite{weininger1988smiles} to encode molecular structure into an ASCII string. Unfortunately, SMILES do not encode information about the spatial organization of amino acids and, therefore, cannot be used for the analysis of loops.}
\rev{To communicate information about amino acids}, multiple tools visualize amino-acid sequences using color-coded lines of letters (\eg NSBI Multiple Sequence Alignment Viewer~\cite{NCBI:MSA}). 
In addition to sequence visualizations, Genome Workbench~\rev{\cite{NCBI:GenomeWorkbench}} also offers a set of integrated tools for studying and analyzing genetic data in general. Unfortunately, it does not provide any means for the analysis of loops. Since it is \rev{common} to compare large numbers of proteins in this way, Nguyen and Ropinski~\cite{nguyen2013large} developed a visualization technique that conveys global patterns in large-scale multiple sequence alignments. However, this technique is also not sufficient for the loops analysis, where higher-level information about the secondary structures and their organization into loops is necessary.

As the analysis of secondary structures plays an important role when studying the function of proteins, several tools emerged that are focusing directly on their visualization. For example, Aquaria~\cite{donoghue2015aquaria} is a web-based tool that enables to map the information about individual secondary structures directly into the 1D representation of protein chains. While Aquaria is utilizing juxtaposition for the comparison of individual proteins, Kocincová et al.~\cite{kocincova2017comparative} proposed a similar solution that uses superimposition instead. Additionally, the authors developed a technique that, to a certain extent, enables the preservation of the spatial information (\eg relative orientation of secondary structures in space) that is embedded in the 1D sequential representation. 
Schulz et al.~\cite{schulz2018uncertainty} investigated how to encode uncertainty of secondary structures assignment into the sequential representations.
While we can borrow some of the concepts from these approaches (e.g., the depiction of secondary structure inside the 1D sequential representation), we need to augment them with additional information (such as flexibility or cross-correlation) before they can be used for the engineering of loops.


\subsection{Visualization of Flexibility and Cross-correlation}
The need to grasp protein dynamics is present in many areas of protein engineering. Thus, several \rev{methods for assessing and visualization of protein flexibility were developed.}
The most common method is using the color or thickness of the 3D representation of a protein to depict so-called B-factors (expressing the fluctuation of individual protein atoms), which can be estimated during the Normal Mode Analysis~\cite{berendsen2000collective}. These visualizations can often be generated by standard tools, such as PyMOL~\cite{delano2002pymol}, or UnityMol~\cite{lv2013game}, assuming that the information about flexibility is available. Usually, a tubular backbone representation is used in these tools since the visual complexity of mapping flexibility on individual atoms would be immense. Alternatively, spatial glyphs (arrows) can provide additional information on the directionality of the movement, as proposed by Bryden et al.~\cite{bryden2010automated}, or show significantly correlated atoms~\cite{tiwari2014webnm}. \rev{However, neither of these methods is suitable for comparison of the flexibility of multiple proteins or considering results from multiple computational approaches. Using superimposition would lead to high occlusion, while juxtaposition of such complex spatial representations leads to separation, which makes it difficult to spot minor differences.}

\rev{Therefore, some tools (e.g., WebNMA~\cite{tiwari2014webnm} or work of Bedoucha et al.~\cite{bedoucha2020visual}) allow the analysis of the correlated motions of individual amino acids using 2D correlation matrices. The matrix axes are composed of protein amino acids, and its cells are colored based on the calculated coupling between these amino acids. Unfortunately, the raw correlation matrices are unsuitable for our particular use case, where we need to depict correlations between loops on two levels (secondary structure correlation and loop correlation) at the same time. Nevertheless, these tools have shown that an abstract visual representation is more suitable for comparative tasks than a 3D view.}



\subsection{Visualization of Protein Loops}
The engineering of protein loops, denoted as loop grafting, is an emerging field. \revca{Nevertheless, apart from the work of our collaborators on the loop grafting pipeline~\cite{schenkmayerova2020engineering, fragrus}, several other groups have active work in the field. For example, Kundert and Kortemme~\cite{kundert2019computational}  presented a computational perspective, including the “closure” of the loop after replacement. Similarly, Mihara et al.~\cite{mihara2021lasso} presented a purely de novo loop design inspired by transplantation techniques. From the experimental point of view, we can list several other works~\cite{ripka2021testing, madden2019exploring, shen2022insights, crean2021loop, feil1998stepwise, smith1995protein}  that are directly related to protein loop transplantation.
}

\revca{However}, \rev{not many visual representations are available to} support the usual workflow of biochemists. To the best of our knowledge, none of the current tools provides the experts with visualization methods beyond 3D views or abstracted representations of secondary structures, as described in the previous sections. 
However, \rev{some tools} focus on \rev{predicting} tertiary structures for loops. 

For instance, SuperLooper2~\cite{ismer2016sl2} is a tool for \rev{predicting} protein segments (\eg loop structures) in globular and helical membrane proteins. The tool is implemented as a derivation of the NGL Viewer\rev{~\cite{rose2018ngl}} and enables the users to find similar protein fragments 
to a selected part of the loaded protein. The best-fitting candidates are presented as a list and, after a manual selection, are also superposed in the 3D View. However, SuperLooper2 leaves many issues unaddressed. 
\rev{For example}, this tool is only suited for tertiary structure prediction and not for rational loop grafting, where \rev{choosing} the Insert protein is important. The tool also does not provide access to measurements important for the loop grafting process, such as the geometry or the flexibility of loops.

Another tool that supports computing and modeling loops is DaReUS-Loop~\cite{karami2019dareus}. 
The results are visualized as a superimposition of (re)modeled loops onto the original protein. The server also offers a table of loops containing useful data, such as the source protein, fitness score, and rigidity of bracing structures. 
However, the tool lacks an interactive link between the important measures and the tertiary structures of the individual loops.
Like the previously discussed tool, DaReUS-Loop is also not suited for rational grafting, where the choice of the Insert protein is important.



\section{Requirements and Task Analysis}
\label{sec:tasks}
\rev{Following the core design study steps~\cite{sedlmair2012design} and principles of nested design model~\cite{munzner2009nested}, we first conducted several informal interviews with protein engineers from a local research group. These discussions provided crucial insights into the challenges of the loop grafting process. 
Furthermore, after analyzing the loop grafting pipeline, we identified several steps where human intervention is necessary and would benefit from visual support. Based on these observations, we formulated the following requirements.} 

\newcommand{\Rnav}{\textbf{R1}}
\newcommand{\Rbook}{\textbf{R2}}
\newcommand{\Rss}{\textbf{R3}}
\newcommand{\RloopsA}{\textbf{R4a}}
\newcommand{\RloopsB}{\textbf{R4b}}
\newcommand{\RloopsC}{\textbf{R4c}}
\newcommand{\Rgraf}{\textbf{R5}}
\newcommand{\Rcolors}{\textbf{R6}}

\rev{
\begin{enumerate}[label=R\arabic*:,leftmargin=0.65cm]
    \item \textbf{Navigation:} The user must always be aware of the current position in the pipeline and understand how to move forward or backward to adjust parameters or fix mistakes.
    \item \textbf{Tracking:} Given the complexity of the pipeline, users need to keep track of the loops they marked as suitable and unsuitable at any point in the pipeline.
    \item \textbf{Secondary Structures:} Since the secondary structures form the loops, the users need access to this information, including the ability to modify the automatic assignment if necessary.
    \item \textbf{Loops:} Users must be able to discern loops suitable for grafting. This process requires information about their geometry~(\RloopsA), flexibility~(\RloopsB), and motion correlation~(\RloopsC).
    \item \textbf{Grafted Proteins:} Ultimately, the users must choose suitable results from modeling tools performing the loop grafting for wet lab testing. This requires information about the 3D structures of the grafted proteins and calculated viability scoring.
    \item \textbf{Color Design:} Finally, we obtained a soft requirement concerning the overall color design choice, as the domain experts desired the final tool to be seamlessly integrated into their existing software portfolio (\url{https://loschmidt.chemi.muni.cz/peg/software}).
\end{enumerate}
}

\rev{Drawing upon the outlined requirements, we proceeded with a specific visualization-oriented task analysis. Using the naming convention from Tamara Munzner's book~\cite{munzner2014visualization} we identified the following \textbf{action$\rightarrow$target} pairs that must be supported.}

\newcommand{\Tide}{\textbf{T1}}
\newcommand{\Tcomp}{\textbf{T2}}
\newcommand{\TcompG}{\textbf{T3}}

\rev{
\begin{enumerate}[label=T\arabic*:,leftmargin=0.65cm]
    \item \textbf{Identify$\rightarrow$Features\&Shape of Loops:} The first part of the grafting pipeline deals with tasks requiring users to identify loops with required (or undesired) properties.
    \item \textbf{Compare$\rightarrow$Features\&Shape of Loops:} The second part of the pipeline requires comparing these properties between multiple loops. This process includes comparing candidate loops from the Scaffold protein with each other and with the loops from the Insert protein.
    \item \textbf{Compare$\rightarrow$Features\&Shape of Grafted Proteins:} Ultimately, users must identify the most suitable grafted proteins for wet lab testing by selecting the best-rated models with the requested features. Nevertheless, the final fine-grain filtering requires comparing individual grafted models with each other, as the highest-scoring model might not be the most suitable for synthesis in the wet lab.
\end{enumerate}
}

\section{\rev{Pipeline Abstraction}}
\rev{Based on the domain-specific requirements and the extracted visualization tasks, we have abstracted the loop grafting pipeline into a six-phase workflow (Section~\ref{sec:workflow}) and designed appropriate visualizations for each phase.} As seen in Figure~\ref{fig:overview}, \rev{we propose to organize these visualizations into four main areas.} The top of the interface is dedicated to the \textit{Workflow Overview} (Figure~\ref{fig:overview}A), informing the users about their progress through the workflow phases (\Rnav). \rev{This progress is intuitively communicated by graying out} incomplete parts of the workflow. \rev{Depicting the entire workflow, instead of displaying only the current phase name, provides users with a sense of completion and informs them about upcoming phases.}

The left part of the interface is dedicated to two views---the \textit{3D View} (Figure~\ref{fig:overview}B, Section~\ref{Sec:3Dview}), showing the spatial arrangement of proteins and loops, and the \textit{Scaffold Loop List} (Figure~\ref{fig:overview}C, \rev{Section~\ref{Sec:ScaffoldList}}), storing candidate loops from Scaffold protein that were already identified by the users (\Rbook). These three views are always visible, as they are relevant for all \rev{workflow phases}.

The right side contains visual support for the individual workflow phases \rev{(Figure~\ref{fig:overview}D, Sections~\ref{Sec:phase1}-\ref{Sec:phase6})}. We opted for a collapsible tab layout, \rev{as it reduces the amount of displayed information, but it still} provides \rev{contextual information} about the previous and upcoming phases when requested by the user. \rev{While an alternative approach would have been to remove other phases entirely, doing so would have deprived users of the ability to revisit and modify individual steps or reference their previous settings.}

\rev{Finally, the overall visual style was designed to ensure our proposed solution looks cohesive with the existing software, emphasizing simplicity and utilizing two main colors on a blue-red scale for better integration (\Rcolors).} \revtwo{We use blue colors for Insert and red for Scaffold proteins, respectively. }


\subsection{Workflow Phases}
\label{sec:workflow}
\rev{We originally considered mirroring the loop grafting pipeline (Section~\ref{sec:background}) step by step. However, upon careful consideration, we decided to group certain pipeline steps into a single visualization phase. We believe that dividing a complex pipeline into smaller phases helps with navigation (\Rnav). Nevertheless, we aim to avoid too much fragmentation so the users would not lose context. Therefore, we designed the workflow such that each phase is now split by an automatic computation step, making it more natural to follow.}

\paragraph*{Phase 1:} 
The first phase corresponds to steps 1-3 of the loop grafting pipeline. \rev{As such, it provides information about the automatic assignment of secondary structures (\Rss) and allows modification if needed. We utilize two bidirectionally linked views. Users can get familiar with the protein shape using 3D View (Figure~\ref{fig:overview}B). But we also designed a simplified 1D View (Figure~\ref{fig:protein_selection}) that is used (with minor modifications) throughout the whole pipeline, informing the user about the secondary structures within the relevant context.}

\paragraph*{Phase 2:} 
\rev{The second} phase is dedicated \rev{to exploring} the results of automatic loop computation and covers steps 4 and 5 of the loop grafting pipeline, focusing primarily on loop geometry \rev{(\RloopsA)}. \rev{To emphasize linking between individual phases and to show loops in the context of the secondary structures (\Rss), we extended the 1D View introduced in the previous phase to encode loop positions with arcs (Figure~\ref{fig:loop-exploration}A). When selecting a loop, users can identify its geometrical features (\Tide) in 3D View as the selected loop is highlighted or perform the initial comparison between the candidate Scaffold and corresponding Insert loops (\Tcomp) using dedicated abstract representations below the 1D View (Figure~\ref{fig:loop-exploration}B and C). 
}

\paragraph*{Phase 3:} 
\rev{To avoid overwhelming users, we separated the flexibility analysis (\RloopsB), which corresponds to steps 6 and 7 of the loop grafting pipeline, into a separate phase. We again utilize the 1D View, now colored by the flexibility of individual secondary structures. Below this view, we provide an abstract representation (Figure \ref{fig:flexibility}A) showing results from different computational algorithms. These visual representations allow users to identify loops (\Tide) with required properties and mark them as candidate loops. Furthermore, depicting all flexibility information at once has the advantage that the users can compare (\Tcomp) the flexibility of the loops and make their decision in a broader context.} 


\paragraph*{Phase 4:} 
In the next phase, \rev{corresponding to steps 8 and 9 of the loop grafting pipeline, we depict the results of cross-correlation analysis between already selected candidate loops from Phases 2 and 3 and all the remaining loops (\RloopsC). We utilize a cross-correlation matrix (Figure \ref{fig:correlation}) known to biologists from NMA Analysis (e.g.,~\cite{tiwari2014webnm}). However, we modified it to allow users to consider both secondary structure correlation and loop correlation simultaneously when identifying remaining possibly interesting candidate loops (\Tide).}

\paragraph*{Phase 5:} 
\rev{The fifth phase, corresponding to step 10 of the loop grafting pipeline, allows the experts to identify pairing between all Scaffold candidate loops selected in Phases 2-4 and Insert loops selected in Phase 2. We provide users with means to attempt to pair the loops automatically. However, manual validation and adjustments are typically needed. Since this pairing is based on the geometrical properties of the loops (\RloopsA), we reuse visual representations from Phase 2 (Figure~\ref{fig:overview}E), allowing users to compare (\Tcomp) the loops. This way, users can work with familiar tools, easing the learning curve.}


\paragraph*{Phase 6:}
\rev{In the} last phase, the paring from the previous phase is used to \rev{model a large amount of} variations of the final grafted protein \rev{in-silico}. \rev{To enable users to identify the best candidates for subsequent in-vitro experiments (\Rgraf), we provide means to display the models in 3D View but also include abstracted 1D views (Figure~\ref{fig:solutions}) suitable for their comparison (\TcompG). This phase corresponds to \rev{the final} steps 11 and 12 of the loop grafting pipeline.}

\section{Design of Individual Visualizations}
\rev{In the following, we describe our rationale behind the design of individual visual representations for each phase in detail.}

\subsection{3D View}
\label{Sec:3Dview}
The 3D View (Figure~\ref{fig:overview}B) displays the 3D conformations of the \rev{Insert and Scaffold} proteins. We utilize the ribbon representation as it emphasizes secondary structures, which play a key role in loop grafting \rev{(\Rss)}. \rev{
Furthermore, this representation is more suitable for the comparison of Insert and Scaffold proteins since it avoids unnecessary visual clutter compared to full atomistic representations.}

The main purpose of the 3D View is to provide visual feedback for manually adjusting the assigned secondary structures (\Rss) and support the identification \rev{(\Tide)} and comparison \rev{(\Tcomp)} of spatial properties of loops and grafted proteins. \revtwo{Nevertheless, we identified several disadvantages of the 3D View. For example, it is problematic to highlight a large amount of loops simultaneously or to compare their biochemical properties. We could not identify a suitable visual channel as shape and position are taken by the secondary structures representation, and color hue is unsuitable for large amounts of categorical data, especially since the loops overlap each other. We attempted to solve this challenge with the dedicated abstract visualizations described below.}

\revtwo{On the other hand, we are losing important spatial information by abstracting from a 3D view. }
\rev{Therefore, the 3D View} is interactively linked with all proposed abstracted representations\rev{, including the Scaffold Loop List.
When a secondary structure or loop is selected, it is automatically highlighted in the 3D View by visually suppressing the rest of the protein with transparency. We also considered using color to highlight selected loops; however, the transparency proved more useful as it better facilitates the comparison of the selected loop from one protein with all the loops from the second protein. This is because the \rev{transparent} protein \rev{with a selected loop} does not obscure the second protein (Figure~\ref{fig:overview}B).}


\subsection{Scaffold Loop List}
\label{Sec:ScaffoldList}
The Scaffold Loop List (Figure~\ref{fig:overview}C) \rev{accompanies all workflow phases as} it provides an overview of all loops of the Scaffold protein \rev{and information about whether they were already evaluated or are yet to be processed by the user (\Rbook)}. Conceptually, we split the Scaffold loops into three possible groups based on the user decisions: \textit{candidate}---loops that are supposed to be grafted, \textit{unsuitable}---\rev{loops marked by the user for removal from exploration, reducing the visual complexity and speeding up the process by avoiding revisiting them}, and \textit{preserved}---loops, where the user has not yet made the final decision about their suitability. At the beginning of the exploration, all loops are marked as preserved. 

The top part of the Scaffold Loop List shows the currently selected candidate loops, while the bottom part displays the preserved Scaffold loops. 
The properties displayed in the list are changing with respect to the current workflow phase. For example, during the Flexibility Evaluation phase, the flexibility scores are displayed, while during the Loop Pairing phase, the list shows the corresponding Insert loops.
The list can be sorted according to those properties, thus providing an overview of the candidate loop properties and support for fast identification of loops with similar properties \rev{(\Tide)} from the preserved Scaffold loops. 
The \faPlusCircle\ and \faMinusCircle\ buttons can be used to move loops between candidate and preserved lists, while the \faTrash\ marks the loop as unsuitable for grafting. 
The \faSearch\ button zooms the 3D View to this particular loop.

\subsection{Phase 1: Visualization of Secondary Structures}
\label{Sec:phase1}

\begin{figure}[htb]
  \centering
  \includegraphics[width=\linewidth]{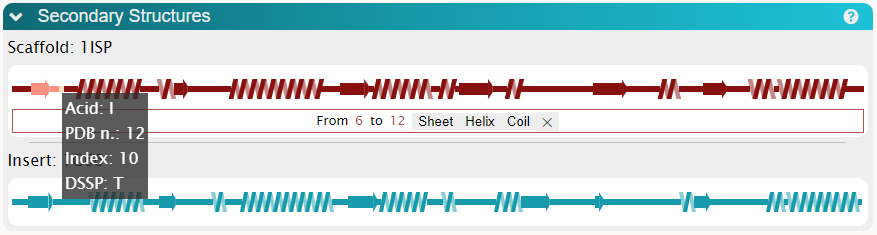}
  \caption{\label{fig:protein_selection}
           The Secondary Structures tab with Scaffold (PDB ID 1isp, red) and Insert (PDB ID 1g66, blue) \rev{proteins. The secondary structures are encoded with glyphs} (helices and arrows). The first \textbeta-sheet from the Scaffold protein and a coil region right after it are brushed \rev{(light red)} to change the secondary structure assignment. The tooltip shows details about the endpoint of the brushed region. }
           
\end{figure}

The first phase of the visualization workflow deals with \rev{exploration of} Scaffold and Insert proteins \rev{and their} secondary structures.
The proteins are depicted in an abstracted 1D representation where the chain is straightened into the 1D line, onto which we encode secondary structures using dedicated glyphs. \rev{We considered several variants of these glyps such as} those used in the Aquaria tool~\cite{donoghue2015aquaria} or by Kocincová et al.~\cite{kocincova2017comparative}. \rev{We decided to \revtwo{encode the two main types of secondary structures using} helices for \textalpha-helices and arrows for \textbeta-sheets (Figure~\ref{fig:protein_selection}) as suggested by Kocincová et al.~\cite{kocincova2017comparative}.}
This encoding is consistent with the ribbon representation employed in the 3D View and enables us to easily add further annotations in the upcoming workflow phases without cluttering the 3D View. 
At the same time, it encodes the crucial characteristics of the protein structures and facilitates the initial evaluation of protein suitability (i.e., sufficient similarity of secondary structures) for loop grafting. \revtwo{We opted to orient the view horizontally to maximize screen space, leveraging the widespread use of monitors with a 16x9 aspect ratio.}

\rev{The 1D View}, along with the 3D View, further facilitate the manual revision of the automatically assigned secondary structures (\Rss). 
This is done by selecting a part of the sequence, either by drawing a selection box over the 1D representation or by clicking at the beginning and the end of the desired region directly in the 3D View. \revtwo{The assignment of the secondary structure to the selected region is performed by clicking on the Sheet, Helix, or Coil buttons in the Secondary Structures tab (see Figure~\ref{fig:protein_selection}).} 
\rev{Following Schneiderman's mantra~\cite{shneiderman1996eyes}, we depict details on demand via} an interactive tooltip on mouse hover \rev{to help the experts set the correct boundaries.} \rev{This tooltip} provides information about individual amino acids, such as the short amino-acid abbreviation, amino acid sequence number, and the current secondary structure assignment (Figure~\ref{fig:protein_selection}). \rev{By hiding the information in the tooltip, we avoid cluttering the view with unnecessary information for all amino acids along the chain.}
After the user performs \rev{the} modification, both 1D and 3D views are updated accordingly. 

\subsection{Phase 2: Visualization of Loop Exploration}
\label{sec:phasetwo}
\begin{figure}[tb]
  \centering
  \includegraphics[width=\linewidth]{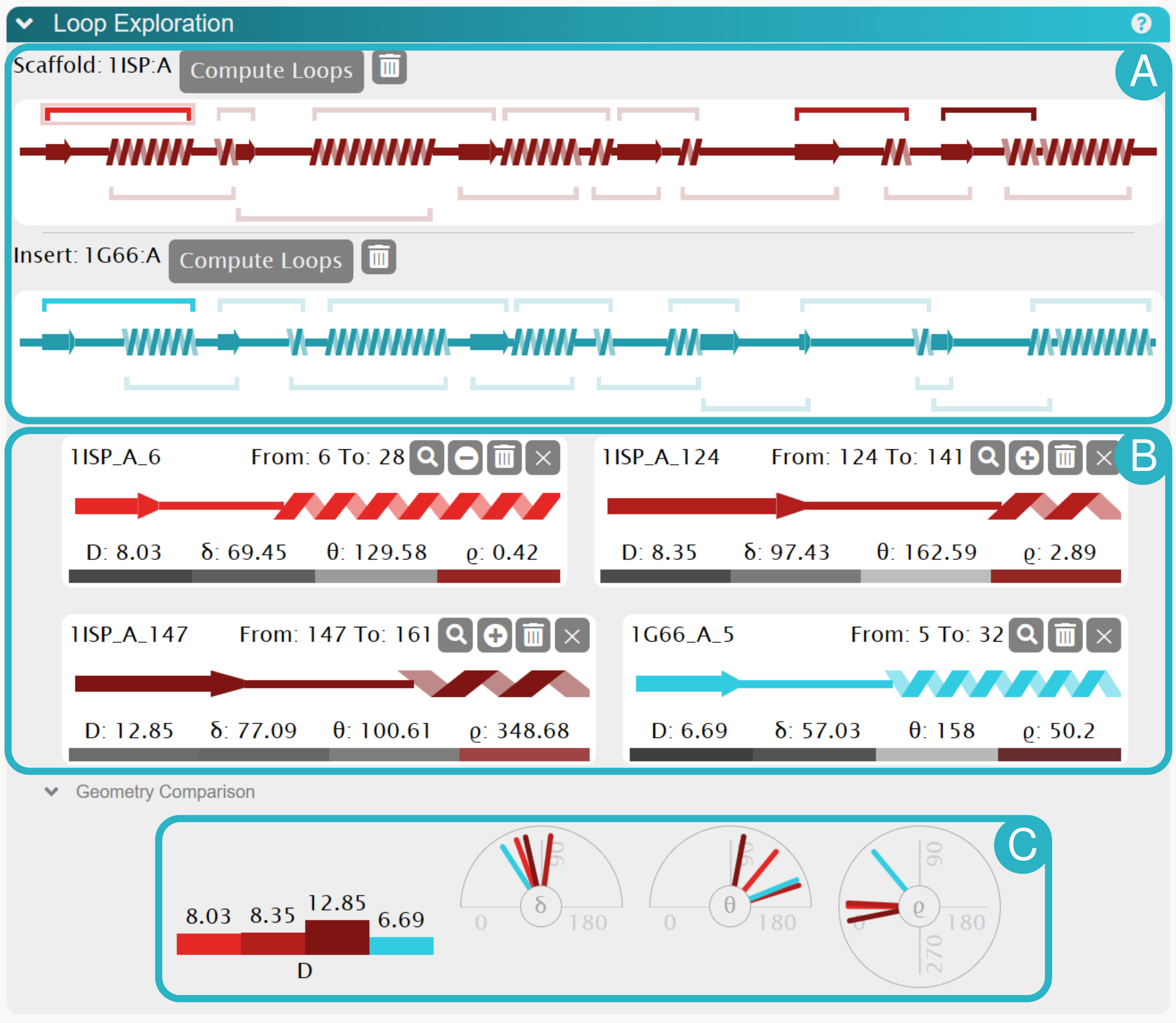}
  \caption{\label{fig:loop-exploration}
           Visual support of the Loop Exploration phase. (A) Overview of the proteins and their loops, (B) selected pairs of loops from the Scaffold and Insert proteins in detail, and (C) depiction of geometric properties of these loops and their comparison.
          }
\end{figure}

The second phase deals with the initial analysis of the loop structures \rev{based on their geometry (\RloopsA)}. 
\revca{Since the loops consist of two periodic secondary structures enclosing one aperiodic region, we needed to resolve a visualization challenge} that the individual loops overlap each other---for a given secondary structure, one loop ends at the structure endpoint, while another starts at its beginning.
\rev{Mapping this information directly into the 3D View, using either color or separate spatial elements, is \rev{im}practical due to the resulting visual clutter. Therefore, we} encoded \rev{loops} into the 1D overview introduced in the previous phase \rev{where we have more screen estate.} Since loops are defined by their starting and ending amino acids, we took inspiration for their encoding from the arc diagrams. \rev{To deal with the overlaps, we} depict the loops as flat bracket-shaped glyphs above and under the 1D representation in an alternating fashion (\rev{Figure~\ref{fig:teaser},} Figure~\ref{fig:loop-exploration}A).
This pattern is broken only when users add custom loops, which can be defined \rev{similarly to} the secondary structures in the previous phase (i.e., by brushing). For these custom loops, we add another layer of glyphs. 


Furthermore, loops tagged as candidates are depicted with a rectangle around them to distinguish already processed loops from unexplored ones (see Figure~\ref{fig:loop-exploration}A).
On hover over the glyph, the loop is highlighted in the 3D View (Figure~\ref{fig:overview}B).
By clicking on \rev{the} glyphs, \rev{the corresponding loops are} selected, \rev{highlighted in 1D overview}, and \rev{can be} explored in the \rev{additional} views below the 1D representations, shown in Figure~\ref{fig:loop-exploration}\rev{B/C}. The purpose of these views is \rev{fast identification of geometric properties of individual loops (\Tide) as well as facilitating comparison (\Tcomp) between the loops.} This is important, especially because the Scaffold loops can be replaced only by Insert loops with similar geometry \cite{fragrus}. \rev{Therefore, we designed} the following two visual representations.



\begin{figure}[htb]
  \centering
  \includegraphics[width=\linewidth]{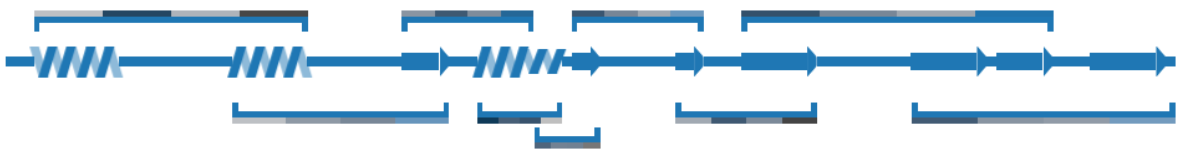}
  \caption{\label{fig:discarded}
           \revca{The original design (later discarded), depicting geometrical properties as grayscale bars directly above the bracket-shaped glyphs representing loops.} 
          }
\end{figure}

\rev{First, each selected loop is depicted in a separate juxtaposed panel using the secondary structures representation (\Rss, Figure~\ref{fig:loop-exploration}B).} \rev{We use the identical hue for the loops as the one assigned to their corresponding proteins, aiming to maintain a clean design and prevent the overuse of colors (\Rcolors).} However, \rev{each loop has a different brightness, generated in a cyclical order from a set of predefined values, to distinguish between the loops from the same protein}. Similarly to the Scaffold Loop List, \rev{each loop can be tagged} as a candidate loop \rev{(\faPlusCircle~button)} or removed from further exploration \rev{(\faTrash~button)}. 
\rev{The grayscale bar} under the 1D loop representation consists of four parts---distance $D$ that represents the Euclidean distance between the endpoints of the coil inside the loop, and three angular $\delta$, $\theta$, and $\rho$ values that capture the spatial arrangements within the loop \cite{archdb, oliva1997automated}. \rev{We were originally considering using a small bar chart, but we decided} to encode these properties with color instead. \rev{While the bar chart would provide a more accurate comparison, using color allows for compact representation, which was more important given the number of loops the users may want to explore.} \rev{Color still supports} identification of \rev{highly mismatched} loop pairs \rev{and if needed, the precise values are depicted in the labels.}
The distance $D$ \rev{and} the angles $\delta$ and $\theta$, ranging from 0° to 180°, are mapped onto a linear grayscale range.
However, angle $\rho$ with angles ranging from 0° to 360° \rev{requires} a special consideration. 
\revca{While} angles around 0° and 360° are \revca{similar, they} would be mapped to very distant values on a linear scale. Therefore, we use circular color scale \includegraphics[width=2.em,height=.8em]{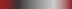} \rev{with} red color for values close to 0 and 360, black for values near 90, and white for values near~270. 
\revca{In one of the first iterations of our design, we also depicted the same color bars above bracket-shaped glyphs representing loops (see Figure~\ref{fig:discarded}). However, domain experts considered this design "too busy and not so important" during the first testing phase.}


The bottom of the Loop Exploration panel shows the \rev{superimposed} representation of the properties of all currently \rev{selected} loops in one place (Figure~\ref{fig:loop-exploration}C) \rev{which is more suitable for comparison} across all loops (\Tcomp). It contains a bar chart for \rev{comparing} distances $D$, where each bar corresponds to one loop, and three angular plots (one for each angle $\delta$, $\theta$, and $\rho$), where loops are depicted as oriented ticks, intuitively encoding the values. We employ color to link the aggregated representations with detailed loop views.

\vspace{3mm}
\subsection{Phase 3: Visualization of Flexibility Evaluation}
\begin{figure}[tb]
  \centering
  \includegraphics[width=\linewidth]{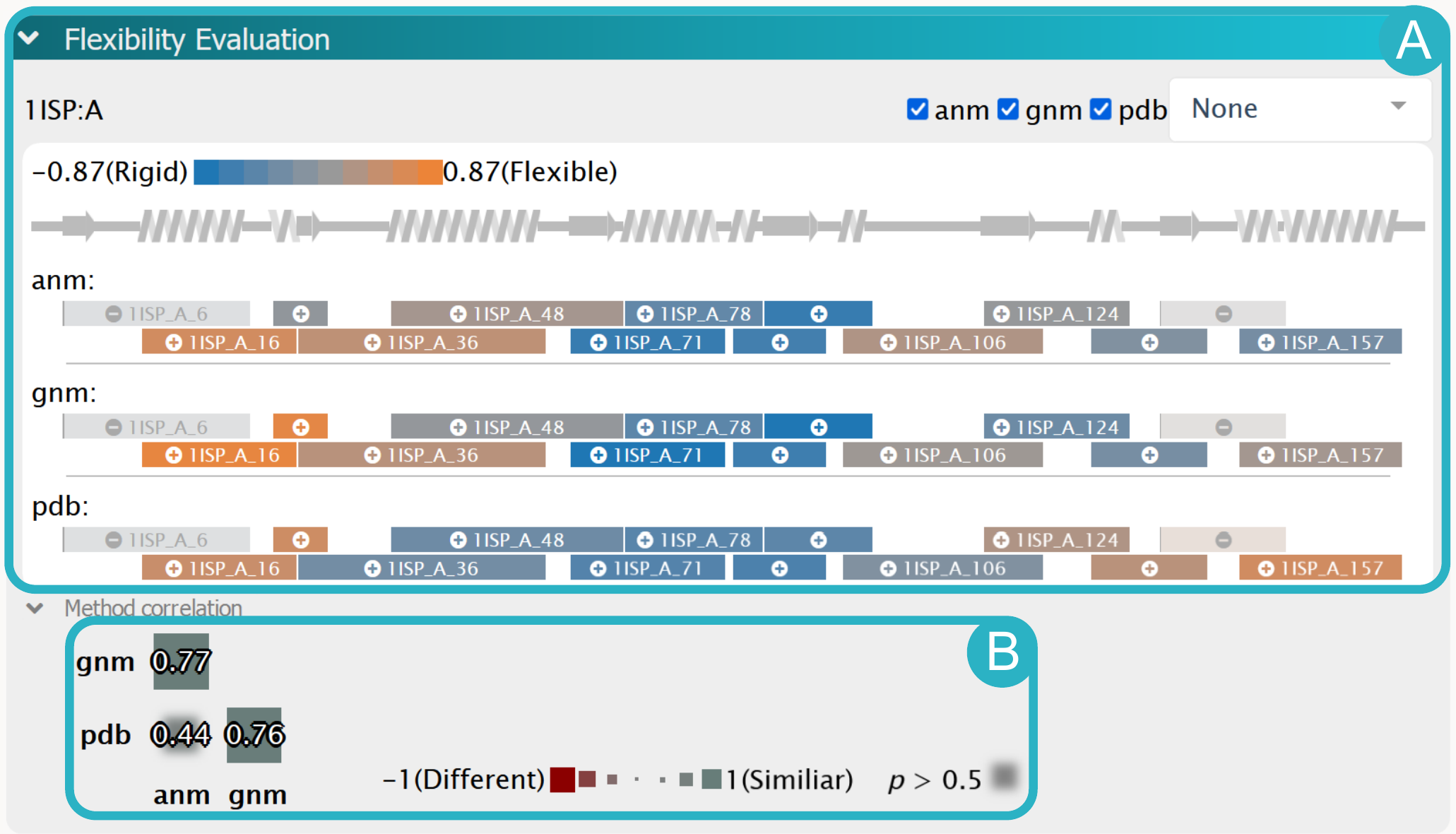}
  \caption{\label{fig:flexibility}
           The Flexibility Evaluation phase showing (A) the flexibility of the Scaffold protein and its loops, detected by three different methods (ANM, GNM, PDB) discussed in the Background section, and (B) the correlation of the values from these three approaches.}
\end{figure}

The third phase is dedicated to the exploration and assessment of the flexibility of the Scaffold protein and its individual loops, further supporting \rev{requirement \RloopsB}.
To support \rev{comparing (\Tcomp)} outcomes of different calculation methods, we juxtapose the loops below the 1D representation of the protein (Figure~\ref{fig:flexibility}A). We use color to encode the flexibility ranging from the most rigid (blue) to the most flexible regions (orange). To improve the interpretability of the color, each loop is \rev{now} depicted as a filled rectangle.

By default, each loop and secondary structure is assigned a color representing its coarse flexibility, i.e., the weighted average flexibility of its amino acids. Here, the users can easily \rev{identify (\Tide)} interesting loops (typically in bright orange) and the differences in the performance of the flexibility prediction methods. 
\rev{When} a more detailed flexibility analysis on the level of amino acids is required, the color coding can be switched accordingly.
The comparison of results of multiple methods serves as validation if a loop is identified by a selected method as highly flexible, but other methods do not support this assumption. Then, the user might reconsider including such a loop in the grafting process.
Loops can be tagged (or untagged) as candidates directly from their rectangular representations \rev{(\faPlusCircle~\faMinusCircle~buttons)}.
Loops \rev{already} selected as candidates are visualized with a lighter color to suppress them and allow users to focus on yet unexplored loops. 

Additionally, to \rev{easily validate} the results of different computational methods, the correlation between them is displayed in a simple correlation matrix (Figure~\ref{fig:flexibility}B). The correlation is expressed as a scalar value that can be either positive or negative. Here, we employ double encoding, with the size showing the absolute correlation value and a diverging color scale depicting positive and negative correlations.
If the correlation of a given method does not reach sufficient significance ($p >$  0.5), it is additionally encoded in the appropriate matrix square by a fuzzy border. \rev{For this, we took inspiration from previous works that use blur to encode the amount of uncertainty and variation~\cite{sterzik2023enhancing, furmanova2020vapor}.} The correlation analysis is an important step since a method with a low correlation with other methods could be an unwanted outlier, and the user might choose to ignore all results from such a method. On the other hand, a negative correlation might suggest a systematic error computation, that can occur when using custom flexibility definitions.

\subsection{Phase 4: Visualization of Correlation Evaluation}
\begin{figure}[htb]
  \centering
  \includegraphics[width=\linewidth]{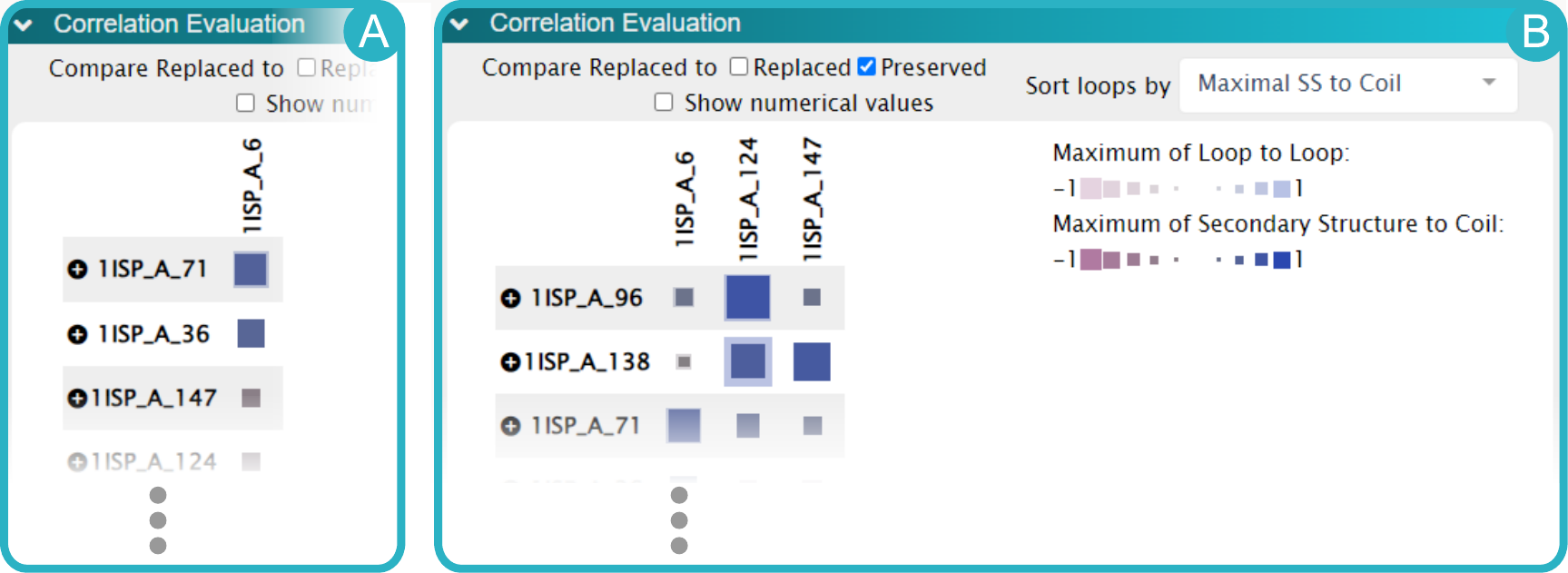}
  \caption{\label{fig:correlation}
           The Correlation Evaluation \rev{phase showing} the correlation of motion between \rev{one (A)} and \rev{three (B)} candidates loops \rev{(columns)} and the remaining preserved loops \rev{(rows)} as colored squares. 
           \rev{In both views, the rows are} \rev{sorted according to motions correlated to the coil section of the loop.}}
\end{figure}

Once an initial set of candidate loops is selected, the additional challenge is \rev{to identify} loops that are expressing highly (positively or negatively) correlated motion with already selected loops and could also be potential candidates \rev{(\RloopsC)}.
This information is presented in a motion-correlation matrix \rev{(Figure~\ref{fig:correlation})}.
\rev{Since there are typically only a few candidate loops, we depict them} in columns and non-candidate loops in rows.
\rev{Neverthelss, c}andidate loops can also be added to rows to \rev{explore} their mutual correlations \rev{if needed}. 
\rev{We} again employ the dual encoding of size and color to identify highly correlated loops and differentiate between positive and negative correlations. Moreover, two motion-correlation properties need to be considered---secondary structure correlation and loop correlation. 
We encode the \rev{secondary structure correlation} by the size of the matrix cell and darker color \rev{scale}. The size of the cell border encodes the loop correlation with a lighter color scale (Figure~\ref{fig:correlation}). 

\rev{We explored the option of using separate representations for the two properties but ultimately decided on a combined view. This choice was made to conserve space and minimize visual decoupling between the properties. \rev{The encoding was also selected such that it} emphasizes secondary structure correlations,} since this property is generally more important for decision-making.
To help the experts \rev{identify} relevant loops \rev{(\Tide)}, we provide an option to sort the loop matrix rows by the correlation values, their position within the protein chain, or their unique identifiers. \rev{When multiple candidate loops are selected, the maximum value on the row is used for sorting.}
The user \rev{can} tag loops as candidates with the \faPlusCircle\ button. 

\subsection{Phase 5: Visualization of Loop Pairing}
In the fifth phase, the suitable Scaffold and Insert loops are selected as input for the grafting algorithm \rev{(\Rgraf)}. 
The users are presented with the same \rev{visual representations (Figure~\ref{fig:overview}E)} that \rev{were} used in the \rev{second} \rev{(}Loop Exploration\rev{)} phase. \rev{The juxtaposed loop panels together with angular charts} are used to \rev{compare geometries (\Tcomp) of} loops from Scaffold and Insert proteins. \rev{This} way, \rev{the user} can verify the viability of replacing the selected loops.


\subsection{Phase 6: Visualization of Loop Grafting}
\label{Sec:phase6}
\begin{figure*}[t]
  \centering
  \vspace{1em}
  \includegraphics[width=\textwidth]{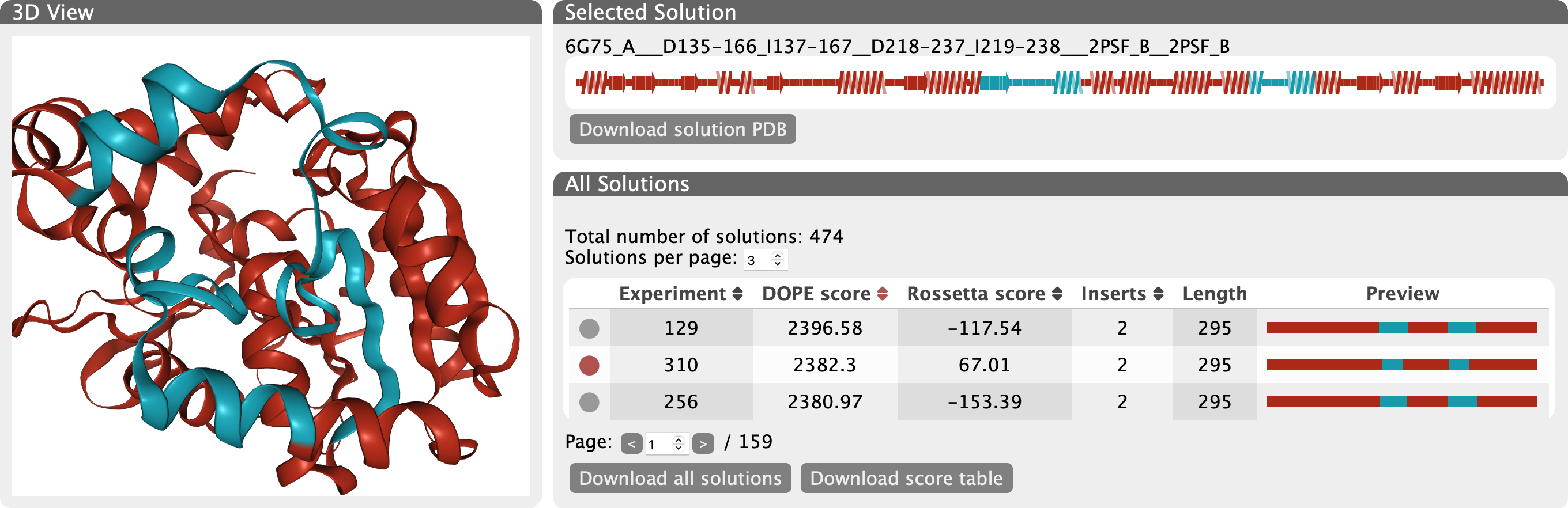}
  \caption{\label{fig:solutions}
             All calculated solutions, with their previews, properties, and scorings, are listed in the final Loop Grafting step. Each solution can be viewed in detail in the 1D representation and 3D View, where the color coding highlights the grafted regions. 
           }
\end{figure*}

Contrary to the previous phases, generating \rev{the grafted protein models} is not a real-time process and may take up to several hours.
Once the grafting process is finished, the user is able to see the list of grafted proteins in a tabular view. Since the computational tools can generate thousands of results, it is important to provide support for \rev{quickly identifying} the most relevant results (\rev{\Tide}). Therefore, the tabular view includes the calculated stability evaluation scores (e.g., coming from DOPE~\cite{shen2006statistical} and Rosetta~\cite{das2008macromolecular}) that can be used for sorting \rev{(the lower the score, the more stable the chimeric protein is predicted to be)}. Additional details about the results, such as the number of grafted loops and the length of the protein chain, are also included.
Furthermore, for each result, the table provides a small graphical preview of the grafted protein, obtained by further abstraction and simplification of the 1D representation. Each protein is depicted as a straight line, and the color is used to encode portions of the original Scaffold protein and the grafted Insert loops. This supports a fast comparison \rev{(\TcompG)} of results, as the differences in loop locations and lengths are clearly visible (Figure~\ref{fig:solutions}).
Individual \rev{selected} solutions can be explored in greater detail in the 3D View.
Here, color is again employed to distinguish between the Scaffold and substituted Insert parts of the solution.

\subsection{Implementation Details}
The proposed interactive visual elements are implemented to be rendered in a browser. The abstract representations are implemented as components in Vue.js JavaScript framework~\cite{you_2014}. \kiraaRemove{and communicates with a stateless REST API implemented as a Python server.} We are leveraging the NGL Viewer~\cite{rose2018ngl} for the 3D View. \kiraaRemove{The tool is distributed as a Docker container. This choice of technologies allows for an environment-agnostic and easy setup in case a third party would be interested in creating a private instance.}
The \rev{original version of LoopGrafter tool (published in NAR~\cite{loopgrafter-nar2022})} is available at \url{https://loschmidt.chemi.muni.cz/loopgrafter}. It includes an interactive guided example, which navigates the users through the loop grafting process and demonstrates the usage of the proposed solution.
The \rev{latest open-source implementation containing the concepts presented in this paper} is available at \url{https://gitlab.fi.muni.cz/visitlab/loopgrafter-frontend-1.2}. 

\section{\rev{Evaluation}}
\rev{Since its release in 2022, the overall utility of our solution is underscored by over \revca{1900} individual grafting jobs performed with the LoopGrafter \cite{loopgrafter-nar2022} tool where \rev{the presented visualizations} have been implemented.}
\rev{Additionally,} LoopGrafter was introduced by one of the paper authors in the 2\textsuperscript{nd} and 3\textsuperscript{rd} hands-on computational enzyme design courses \cite{academia} to 60 biochemical experts in each edition. The participants \revtwo{had varying levels of expertise in protein engineering.  They} were \rev{encouraged to follow a guided example} and to try their own researched proteins. The vast majority of the experts who took part in the courses expressed satisfaction with the \rev{visual representations} and agreed that \rev{they make} the grafting process more accessible, and 
helped them understand the process. 
\rev{Despite this positive feedback,} four participants had difficulties using the tool. \rev{The informal interviews with these four participants indicated an overall lack of familiarity with the loop grafting concept. This insight highlights a potential area for future work aimed at enhancing accessibility for complete novices.} 

\rev{Prior to its release, the} LoopGrafter application was tested and evaluated on 50 \revca{homologous protein pairs to ensure the proposed pipeline produces meaningful results. During this testing phase, domain experts experienced the proposed visualizations firsthand and evaluated their suitability for the grafting pipeline. As a result, one of the visual representations was discarded (see Section~\ref{sec:phasetwo} and Figure~\ref{fig:discarded}). Additionally, minor adjustments to the color coding of the motion-correlation matrix were made to better adhere to the conventions used in their field. 

While the experts evaluated the feasibility of the best-predicted results for each of the 50 cases, they were not tested experimentally. Therefore, the solution was thoroughly evaluated on additional 5 case studies replicating results from the literature. This evaluation was performed} by two senior researchers with expertise in protein engineering, who \rev{contributed significantly to the design of visual interactive elements and co-authored} this paper. \rev{To} illustrate the benefits of the proposed visualizations, we recall here \rev{one such case where one of the experts attempted to replicate }a loop-grafting \revca{effort~\cite{bacterium2008}. Namely} the replacement of a loop in protein lipase A in \emph{B. subtilis} (PDB ID 1isp) with its structurally equivalent loop from protein acetylxylan esterase in \emph{P. purpurogenum} (PDB ID 1g66). 

After downloading \rev{the structures} from the PDB \cite{bermanPDB2000} and automatically calculating their secondary structure\rev{s}, the 1D representation of proteins in the first phase (Figure~\ref{fig:protein_selection}) facilitat\rev{ed} checking the assigned secondary structure for potential artifacts. 
The authors of the original study directly mention that the first \mbox{\textbeta-strand} element could be extended. 
\rev{Thus, the expert} reassigned residues 10-12 in the lipase and 11-13 in the esterase to \rev{\textbeta-}strand conformation \rev{by clicking and dragging in the desired region on the 1D representation of the protein. To achieve similar results without the visual support, a separate file with the DSSP assignment would need to be manually edited for each protein, making the process more tedious.}

Also, \rev{the expert} observed that 
the first 4 (lipase) or 3 (esterase) residues of the subsequent helical element are in 3:10 \rev{conformation (marked in the tooltip as DSSP:G)} instead of \textalpha~\rev{(noted as H)}. \rev{Since 3:10 conformations are less regular and more dynamic than \textalpha-helices, the experts converted these regions to “coil”.} This observation improved the overall grafting in later phases. \rev{Considering that this feature was ignored in the original study, it exemplifies the effectiveness of our solution in identifying residues that may benefit from an alternative secondary structure assignment.}  
The 3D View and detailed secondary structure assignment obtained by hovering on the 1D view were especially highlighted as useful by the experts. \rev{An alternative would imply working simultaneously with three different tools. One for 3D visualization, another with the fine secondary structure assignment from DSSP, and finally, an annotation of the extent of the loop. Our design integrates all the information in one place, making the decision-making process easier.}

In the Loop Exploration phase, all calculated loops were depicted in the 1D representation, allowing on-click actions \rev{to examine and compare} their geometrical properties.
From the original publication, we know that residues 12-20 from the lipase were substituted by the equivalent residues 12-23 from the esterase. Looking at the loops that matched these corresponding positions, the loops 1ISP\_A\_6 from the Scaffold and 1G66\_A\_5 from the Insert were identified \rev{by the expert} as those to take part in the replacement. 
\rev{Using} \rev{glyphs and the angular plots (Figure~\ref{fig:loop-exploration}), the expert evaluated} the geometrical differences between the two loops. They differ by approx. 1.6 {\AA} in the distance D and 19\textdegree, 15\textdegree, and 26\textdegree\ in the angles \textdelta, \texttheta, and \textrho, respectively. \rev{An alternative approach would consist of listing and (manually) comparing the different angles. Thus, the visual support provides a much more intuitive access to the similarity between loop geometries.} \rev{Note that t}hese values are generally lower than the ones that would be obtained if the modifications on the secondary structure assignment \rev{from} the previous phase were not applied (D$\approx$1.3 \AA, \textdelta$\approx$22\textdegree, \texttheta$\approx$38\textdegree, \textrho$\approx$50\textdegree). \rev{The expert, therefore,} added \rev{1ISP\_A\_6} loop to the list of candidates.


\rev{In the next step, the expert proceeded with assessing the} flexibility of loops \rev{and secondary structures} in the Scaffold (Figure~\ref{fig:flexibility}). 
Looking at any of the three \rev{computation} methods available, it was clear that the selected loop lies in the boundary of one of the most flexible areas of the input proteins. This property \rev{often indicates} a region in the enzyme that has an impact in making the substrate available to the catalytic machinery, which confirms the usefulness of the loop as a grafting target~\cite{TOTHPETROCZY2014131}. \rev{Additionally,} the correlation matrix showing the results of \rev{all} flexibility calculation methods indicat\rev{ed} that none yield out-of-margin outcomes and that there is a reasonable agreement between them. This observation \rev{assured the expert} of choosing the correct candidate loop for further steps. 

The subsequent Correlation Evaluation phase seems irrelevant in this use case since the authors of the original publication only transplanted a single loop. However, \rev{we can illustrate the benefits of our approach and showcase} what could have been learned from this analysis. Among the different metrics offered for sorting, “Maximal SS to Coil” \rev{could have been} used, as this metric is illustrative of the motions correlated to the coil section of the loop, \rev{indicating} substrate recognition and activity. Here, loops 1ISP\_A\_71 \rev{and} 1ISP\_A\_36 stand out with the highest positive cross-correlations \rev{(prominent blue rectangles in Figure~\ref{fig:correlation}A)}, and loops 1ISP\_A\_147 and 1ISP\_A\_124 with the most negative ones \rev{(as smaller rectangles on the red scale)}. 1ISP\_A\_36 is not a surprising outcome of this analysis since this loop is the structural neighbor to the selected one. They share a close vicinity \rev{to} the central \textbeta-sheet of the fold and are expected to have correlated motions. 

More interesting are the other three top-ranked loops from this analysis. The first one, 1ISP\_A\_71, contains the key catalytic residue, serine 77 in the lipase. The fact that this loop contains an important residue makes it unsuitable for grafting unless the same residue in a similar 3D position and conformation will be present in the replacement. The other two loops \rev{(1ISP\_A\_147 and 1ISP\_A\_124)} have indeed been targets of experimental mutational engineering to improve the thermostability and catalytic efficiency of the enzyme \rev{in a different study} \cite{Singh2015-eq}. Thus, the utility of this analysis phase: upon previous knowledge about the importance of the selected loop (1ISP\_A\_6) for the stereoselectivity of the enzyme, three other regions emerge that are either key to its activity (1ISP\_A\_71) or have been targeted to improve the performance of the enzyme (1ISP\_A\_124 and 1ISP\_A\_147). These two latter loops would have been good targets along with 1ISP\_A\_6. 
This \rev{confirms the ability of our solution to} visually identify potentially important loops and, therefore, speed up the experiment design. 

\rev{In the Loop Pairing phase, the expert easily} matched the Scaffold loop candidates to their corresponding Insert equivalents by intuitive point-and-click interaction \rev{(Figure~\ref{fig:overview})}. To this end, the 1D representations of both proteins \rev{were} used. By checking the angular plots, \rev{the expert was able to assess the geometric properties of individual loop pairs} to verify no significant deviations between them. \rev{An alternative approach would again consist of listing and (manually) comparing the different angles and distances. 
} 

\rev{In the Loop Grafting phase,} the Scaffold and Insert loop pairs \rev{were} submitted for actual grafting. 
Once the results \rev{were} ready, \rev{the experts} browsed \rev{the} list of 1D representations of the grafted proteins 
accompanied by relevant metrics that indicate the viability of the grafted construct. \rev{In this way, the experts were able to obtain an energetic assessment, and at the same time, they could evaluate how the chimeric protein looks.}
In our case study, the best solutions suggest grafting a longer region, between residues 9 and 26, instead of the 12-20 range chosen by the authors of the original study.
\rev{While we did not validate the results by experiments, the expert suspected that if the authors of the original study tried the longer region suggested by our solution, they would likely find a more stable protein.}

\rev{After the experts tested the LoopGrafter on the aforementioned 50 case studies, we asked them to summarize their} point of view. \rev{They mentioned} the proposed visualizations help\rev{ed them} understand the properties of individual loops \rev{and} design their feasible replacements. \rev{They confirmed that the visual support can} significantly reduce the number of shoot-and-miss experiments. \rev{The experts appreciated the ability to change definitions of secondary structures at will.}
Another benefit \rev{mentioned by the experts was} that the visualizations make \rev{identifying} the dynamic role of each protein secondary structure element easier, allowing for spotting loop grafting candidates. \rev{They also considered} the layout useful for spotting equivalent loops in the Insert protein, from which fragments \rev{should} be transferred to the Scaffold protein. \rev{Both bar chart and angular plots} greatly help in this regard \rev{as they provide an intuitive comparison between the geometries of the loops to be exchanged.} Finally, \rev{the experts mentioned that} the \rev{visualization of} grafted proteins \rev{helped them decide which models are suitable for} further experimentation in the laboratory. \rev{Nevertheless, the experts also mentioned that while useful for overview purposes, the size of 3D View hindered their analysis, and they wished for a more powerful 3D viewer with the ability to full-screen mode.} 
\rev{The experts also pointed out that some of the most advanced features, such as the secondary structure re-definition, are not immediately apparent, although they are easy to understand once properly explained.}

\section{\revca{Discussion and Limitations}}
Based on our experience with the design of \rev{our} solution and received feedback, we formulated several design recommendations \revca{that can serve the broader visualization community. Firstly, we would like to highlight the importance of the iterative design process in the search for representations adhering to the conventions and needs of the application domain. Such conventions (e.g., colors or commonly used encodings within the domain) might necessitate compromises or adjustments in visualization design, but considering them during the design process leads to intuitive and easily adoptable solutions for the domain experts. Further, }
our findings emphasized that grouping semantically related steps of a complex pipeline into discrete phases, depicted in collapsible tabs, significantly aids user navigation. \revca{This design} also avoids unnecessary fragmentation of the tasks and preserves contextual information within individual phases. 
In our case, the abstraction process led to several comparative tasks present throughout the entire workflow. Such tasks are commonly occurring in many application areas dealing with spatial data analysis. We found that employing common visual metaphors and reusing visual representations across different phases seems to create useful links for transferring knowledge between the phases and shorten the learning curve \revca{when it comes to the meaning of visual representations}. Finally, while 2D abstract representations are useful for comparison purposes, they cannot fully replace spatial views, emphasizing the importance of bidirectional linked views.

Although the conducted expert evaluation confirmed that currently implemented interactive elements provide helpful support for the loop grafting pipeline, \revtwo{it also revealed several limitations, including the need for better handling of 3D visualization and enhancing accessibility of advanced features for novices. Another possible limitation is the scalability of 1D representation. While our solution supports \revca{small to medium-sized} proteins that are currently typically used in loop grafting, for proteins with much longer sequences the representation might need to be modified. A simple adjustment would be to enable horizontal scrolling through the protein chain. An alternative solution would be aggregating the chain to fit the screen width and enable zooming. However, here we would have to find an appropriate aggregation strategy.}

\section{Conclusion and Future Work}

In this paper, we presented interactive visual elements that follow our proposed visualization workflow for protein loop grafting. 
Within the presented case study, we demonstrated the performance of the presented visual support applied to \rev{real a scenario}.



As the modular architecture of the tool enables the simple addition of new steps to the pipeline, we are ready to address the ongoing research efforts in the design of the pipeline and its enhancements.
\revca{While our tool was designed to replace loops consisting of two periodic secondary structures flanking one aperiodic region, it may be used for other purposes. Through the re-definition of secondary structures, users can attempt to replace larger portions of the protein (even domains) or, on the contrary, only the flexible coils.}

We are also planning to extend visualizations to support more general use cases, such as the exploration and detection of important regions in a single protein, and to test their replacement by geometrically equivalent regions from multiple different proteins \rev{simultaneously}. For~that, we are planning to further collaborate with domain experts on the development of visual support for the tool that builds on top of methods described in \mbox{Bonet et al.~\cite{fragrus}}.

\acknowledgments{%
The presented work has been supported by the Technology Agency of Czech Republic project no. FW03010208, by the Czech Ministry of Education project Elixir CZ no. LM2023055, the National Institute for Cancer Research project no. LX22NPO5102, and the National Institute for Neurology Research project no. LX22NPO5107, financed by European Union - Next Generation EU. 
Parts of this work were also conducted within the context of the Center for Data Science (CEDAS) at the University of Bergen and supported by the University of Bergen and the Trond Mohn Foundation in Bergen (\#813558, Visualizing Data Science for Large Scale Hypothesis Management in Imaging Biomarker Discovery (VIDI)).}

\bibliographystyle{abbrv-doi-hyperref}

\bibliography{template}

\end{document}